\documentclass[letter]{aa}
\usepackage[english]{babel}  
\usepackage{graphicx}  
\usepackage[utf8]{inputenc}  
\usepackage{microtype}  
\usepackage{booktabs}  
\usepackage{rotating}  
\usepackage{lscape}
\usepackage[squaren]{SIunits}
\usepackage{natbib}
\usepackage{amsmath}
\usepackage{amssymb}
\usepackage{pstricks}
\usepackage{txfonts}
\usepackage{tikz}
\usepackage{color}
\usepackage{hyperref}

\newcommand{\tex}{T_\mathrm{{ex}}}
\newcommand{\tmb}{T_\mathrm{{MB}}}

\newcommand{\td}{T_\mathrm{{d}}}

\newcommand{\pot}[1]{10^{#1}}

\newcommand{\vlsr}{V_\mathrm{{LSR}}}

\newcommand{\cm}{\usk\centi \metre}

\newcommand{\pc}{\usk\mathrm{pc}}

\newcommand{\kpc}{\usk\mathrm{kpc}}

\newcommand{\msun}{\usk\mathrm{M_\odot}} 
\newcommand{\msuntab}{\mathrm{M_\odot}} 
\newcommand{\jy}{\usk\mathrm{Jy}}

\newcommand{\yr}{\usk\mathrm{yr}}

\newcommand{\lsuntab}{\mathrm{L_\odot}}

\newcommand{\beam}{\usk\mathrm{beam}}

\newcommand{\kms}{\usk\kilo\metre\usk\second^{-1}}
\newcommand{\kmstab}{\kilo\metre\usk\second^{-1}}
\newcommand{\kel}{\usk\kelvin}
\newcommand{\mum}{\usk\micro\metre}

\newcommand{\vs}{\ensuremath{\mathrm{\,vs.\,}}}
\newcommand{\citepSabatini}{(Sabatini et al., in prep.)}
\newcommand{\citealtSabatini}{Sabatini et al., in prep.}
\newcommand{\citepLeurini}{(Leurini et al., subm.)}
\newcommand{\citealtLeurini}{Leurini et al., subm.}

\newcommand{\hhdp}{\ensuremath{\mathrm{H}_{2}\mathrm{D}^{+}}}
\newcommand{\nndp}{\ensuremath{\mathrm{N}_{2}\mathrm{D}^{+}}}
\newcommand{\mol}[1]{\ensuremath{\mathrm{#1}}}

\title{A timeline for massive star-forming regions via combined observation of o-\hhdp\ and \nndp}
\author{
    A. Giannetti \inst{\ref{ira}}
    \and S. Bovino \inst{\ref{conce}}
    \and P. Caselli \inst{\ref{mpe}}
    \and S. Leurini \inst{\ref{oac},\ref{mpi}}
    \and D. R. G. Schleicher \inst{\ref{conce}}
    \and B. K\"ortgen \inst{\ref{hamburg}}
    \and K.~M. Menten \inst{\ref{mpi}}
    \and T. Pillai \inst{\ref{boston}}
    \and F. Wyrowski \inst{\ref{mpi}} 
}
\institute{
    INAF - Istituto di Radioastronomia \& Italian ALMA Regional Centre, Via P. Gobetti 101, I-40129 Bologna, Italy\label{ira}
    \and Departamento de Astronom\'ia, Universidad de Concepción, Barrio Universitario, Concepción, Chile \label{conce}
    \and Centre for Astrochemical Studies, Max-Planck-Institute for Extraterrestrial Physics, Giessenbachstrasse 1, 85748 Garching, Germany\label{mpe}
    \and INAF-Osservatorio Astronomico di Cagliari, Via della Scienza 5, I-09047, Selargius (CA), Italy \label{oac}
    \and Max-Planck-Institut f\"ur Radioastronomie, auf dem H\"ugel 69, D-53121, Bonn, Germany \label{mpi}
    \and Hamburger Sternwarte, Universit\"at Hamburg, Gojenbergsweg 112, D-21029 Hamburg, Germany \label{hamburg}
    \and Institute for Astrophysical Research, Boston University, 725 Commonwealth Ave, Boston, MA 02215 \label{boston}
}

\abstract{
    In cold and dense gas prior to the formation of young stellar objects, heavy molecular species (including CO) are accreted onto dust grains. Under these conditions \mol{H_3^+} and its deuterated isotopologues become more abundant, enhancing the  deuterium fraction of molecules such as \mol{N_2H^+} that are formed via ion-neutral reactions. Because this process is extremely temperature sensitive, the abundance of these species is likely linked to the evolutionary stage of the source.
}
{
    We investigate how the abundances of o-\hhdp\ and \nndp\ vary with evolution in high-mass clumps.
}
{
    We observed with APEX the ground-state transitions of o-\hhdp\ near $372\usk\giga\hertz$, and \nndp(3--2) near $231\usk\giga\hertz$ for three massive clumps in different evolutionary stages. The sources were selected within the G351.77--0.51 complex to minimise the variation of initial chemical conditions, and to remove distance effects. We modelled their dust continuum emission to estimate their physical properties, and also modelled their spectra under the assumption of local thermodynamic equilibrium to calculate beam-averaged abundances.
}
{
    We find an anticorrelation between the abundance of o-\hhdp\ and that of \nndp, with the former decreasing and the latter increasing with evolution. With the new observations we are also able to provide a qualitative upper limit to the age of the youngest clump of about 10$^5$ yr, comparable to its current free-fall time.
}
{
    We can explain the evolution of the two tracers with simple considerations on the chemical formation paths, depletion of heavy elements, and evaporation from the grains. We therefore propose that the joint observation and the relative abundance of o-\hhdp\ and \nndp\ can act as an efficient tracer of the evolutionary stages of the star-formation process.
}

\keywords{stars: formation, ISM: abundances, ISM: molecules}
\offprints{A. Giannetti (agianne@ira.inaf.it)}                                                  

\begin{document}  
    
    \maketitle
    
    \section{Introduction}\label{sec:intro}
    
    Depletion of heavy elements proceeds progressively in dense and cold material ($n_{\mol{H_{2}}} > \mathrm{few} \times \pot{4}\cm^{-3}$, $T\lesssim25\kel$), rendering heavy molecular species unobservable, and thus not viable to trace gas in the earliest phases of the star-formation process. 
    It has been demonstrated that deuterated species become more abundant under these conditions,
    and that there is a correlation between the degree of deuteration and the depletion of CO \citep{Caselli+99_apjl523_165, Caselli+02_apj565_344, Bacmann+03_apjl585_55}. Large values for the deuterium fraction are also found towards protostellar sources \citep[e.g.][]{Ceccarelli+98_aap338_43, Parise+02_aap393_49}.

    Pivotal to the deuteration process is the exothermic reaction chain $\mol{H_{3}^{+}} \rightarrow \mol{H_{2}D^{+}} \rightarrow \mol{D_{2}H^{+}} \rightarrow \mol{D_{3}^{+}}$ \citep[e.g.][]{Walmsley+04_aap418_1035, Flower+04_aap427_887}.
    \hhdp, \mol{D_{2}H^{+}}, and \mol{D_{3}^{+}} react with other species, producing the vastly enhanced deuterium fractions in molecules at the core of the ion-neutral chemistry. \mol{N_{2}} in particular can react with \hhdp\ to form \nndp via
    \begin{equation}
    \mol{N_{2}} + \hhdp \rightarrow \nndp + \mol{H_{2}},\label{react:nndp_formation}
    \end{equation}
    and more efficiently with \mol{D_2H^+} and \mol{D_3^+}. The molecule
    \nndp\ is frequently used as a reliable tracer for gas affected by heavy CO depletion \citep[e.g.][]{Caselli+02_apj565_344, Barnes+16_mnras458_1990}.
    Because \hhdp\ and \nndp\ are chemically linked, and because they are abundant only under specific circumstances, they could be used to determine the evolutionary stage of a source in early phases \citep[e.g.][]{Caselli+08_aap492_703, Emprechtinger+09_aap493_89, Fontani+15_aap575_87}, something of particular interest in the high-mass regime.
    
    \citet{Pillai+12_apj751_135} have obtained maps for part of the DR21 complex of the o-\hhdp$(1_{1,0} - 1_{1,1})$ 
    and \nndp(3--2) transitions, observed with the James-Clerk-Maxwell Telescope (JCMT) and the Submillimeter Array (SMA), respectively. They find very extended o-\hhdp\ emission as do \citet{Vastel+06_apj645_1198} in the low-mass source L1544. They also find that this species mainly traces gas that is not seen in dust continuum emission or in the interferometric \nndp\ data. \hhdp\ may therefore be sensitive to gas that would elude detection in the most commonly used tracers, and can represent an even earlier stage in the process of star formation. 
    
    These results raise two important questions that we try to address in this work: Firstly, is the lack of correlation between o-\hhdp$(1_{1,0} - 1_{1,1})$ and \nndp(3--2) found by \citet{Pillai+12_apj751_135} real, or is it an effect of interferometric filtering? Secondly, how do the abundances of these molecules evolve with time, and why? In this Letter we first describe our observations, present the results, and finally discuss the importance of our findings.
    
    \section{Source selection and observations}\label{sec:obs_and_sample}
    
\begin{figure}
        \includegraphics[width=\columnwidth]{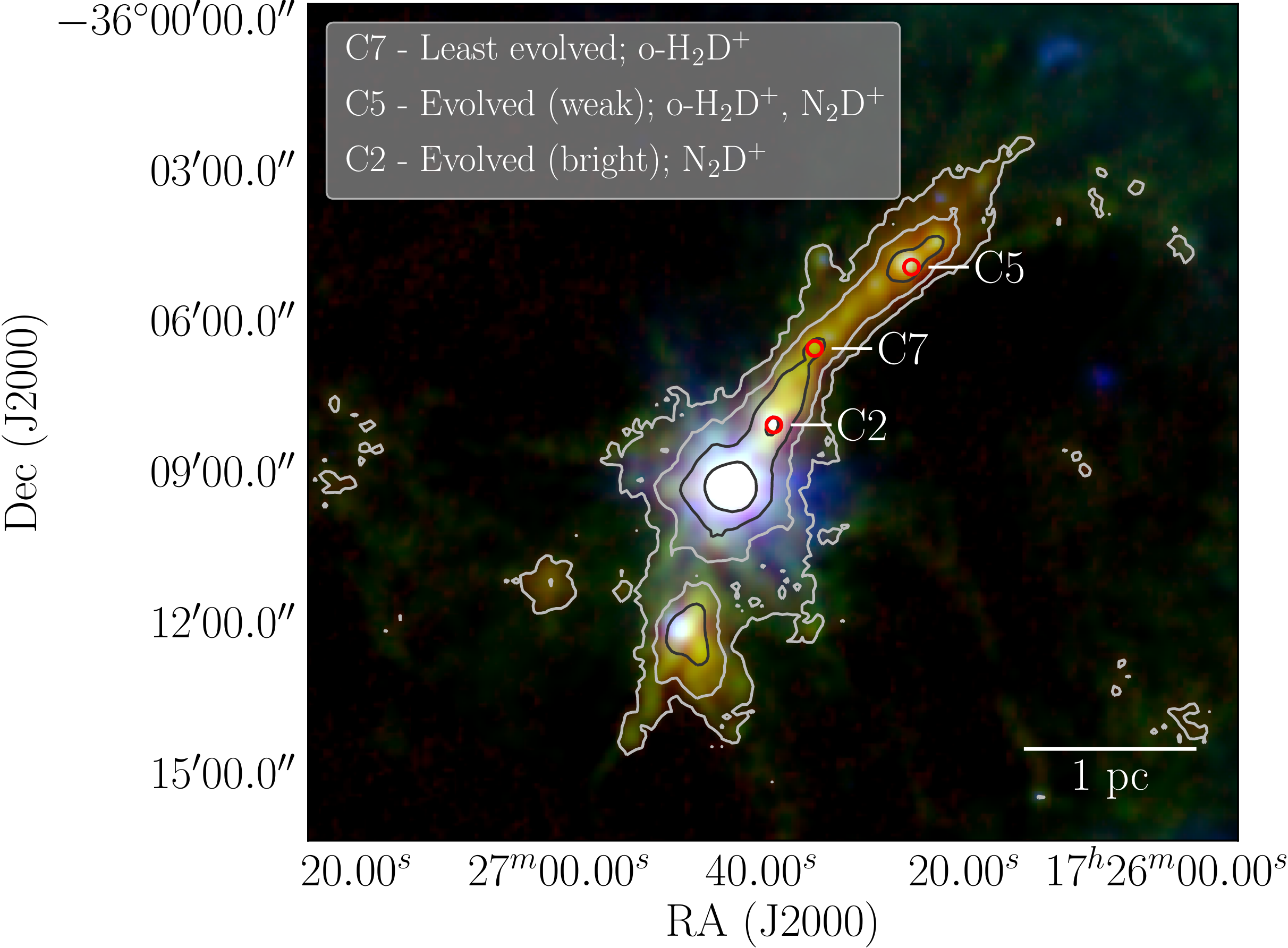}
        \caption{Three-colour image (red: ATLASGAL $870\mum$, green: Hi-GAL $250\mum$, blue: Hi-GAL $70\mum$) of the G351 complex. Contours of the ATLASGAL images are also indicated in grey ($0.15, 0.5, 1.5, 5\usk\jy\usk\beam^{-1}$). Clumps 2, 5, and 7 are indicated in red. In the top-left corner we list the clumps from the least- to the most evolved, and the species we observed in each source.}\label{fig:multilambda}
    \end{figure}

    G351.77--0.51 (hereafter G351) is the closest and most massive filament ($D=1\kpc$, $M\sim2000\msun$, \citealtLeurini) identified in the $870\mum$ APEX Telescope Large Survey of the Galaxy (ATLASGAL) \citep{Schuller+09_aap504_415}. 
    A three colour image of the region is shown in Fig.~\ref{fig:multilambda}, emphasising the large amount of dense and cold gas in its massive ridge.
    
    G351 offers several advantages for our purposes: 1) it is close-by, 2) it hosts massive clumps that exceed the \citet{KauffmannPillai10_apjl723_7} threshold for high-mass star formation, and 3) the clumps are in different evolutionary stages. Therefore, sources along the spine of the G351 filament are extremely well suited to studying the evolution of chemistry in high-mass clumps, not only because they are nearby and all at the same distance, but also because they have similar properties and share the same initial chemical conditions.     
    
    We select three clumps in this complex \citep[Clumps 2, 5, and 7, following ][]{Leurini+11_aap533_85} in different evolutionary stages, based on their IR properties:
    Clump 7 is the least evolved, being still quiescent at $70\mum$; Clumps 5 and 2 are both bright at this wavelength, with the former being weaker than the latter (Fig.~\ref{fig:multilambda}).
    These sources are the ones with the largest column densities and masses found in G351 in each evolutionary stage. Their mass and peak column densities are also within a factor of two, using a temperature of $25\kel$ for Clump 2, and $10\kel$ for Clumps 5 and 7 (cf. Table~4 in \citealt{Leurini+11_aap533_85}).
    
    The dust emission peak for each of the three clumps has been targeted by APEX 12m submillimeter telescope observations (project M-099.F-9508A), using the PI230 and FLASH$^+$ receivers, to cover the \nndp(3--2) and o-\hhdp$(1_{1,0} - 1_{1,1})$ lines at $231.3216$ and $372.42134\usk\giga\hertz$, respectively.
    Observations were performed between 2017 May 31 and 2017 September 19, and completed on 2018 July 1-2. 
    The rms noise on the main-beam brightness temperature scale, $\tmb$, is $\sim20\usk\milli\kelvin$ at $231\usk\giga\hertz$ and $30\usk\milli\kelvin$ at $372\usk\giga\hertz$, for a spectral resolution of $0.45\kms$. 
    We converted the antenna temperature $T_A^*$ to $\tmb$, using $\eta_{\mathrm{MB}} = 0.60$ for o-\hhdp$(1_{1,0} - 1_{1,1})$ and $\eta_{\mathrm{MB}} = 0.69$ for \nndp(3--2).
    
    \section{Results}\label{sec:results}
    
\begin{figure*}
        \includegraphics[width=0.5\textwidth]{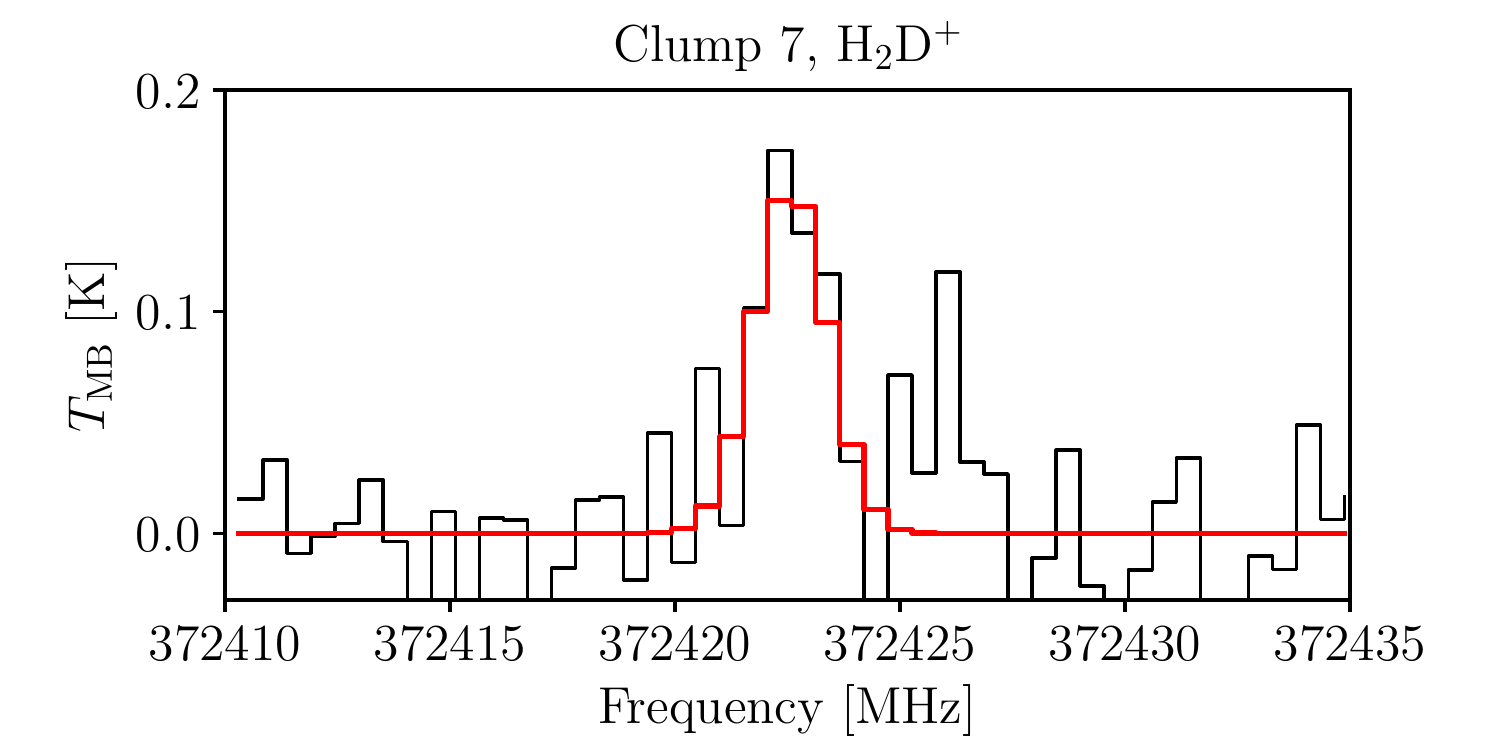}
        \includegraphics[width=0.5\textwidth]{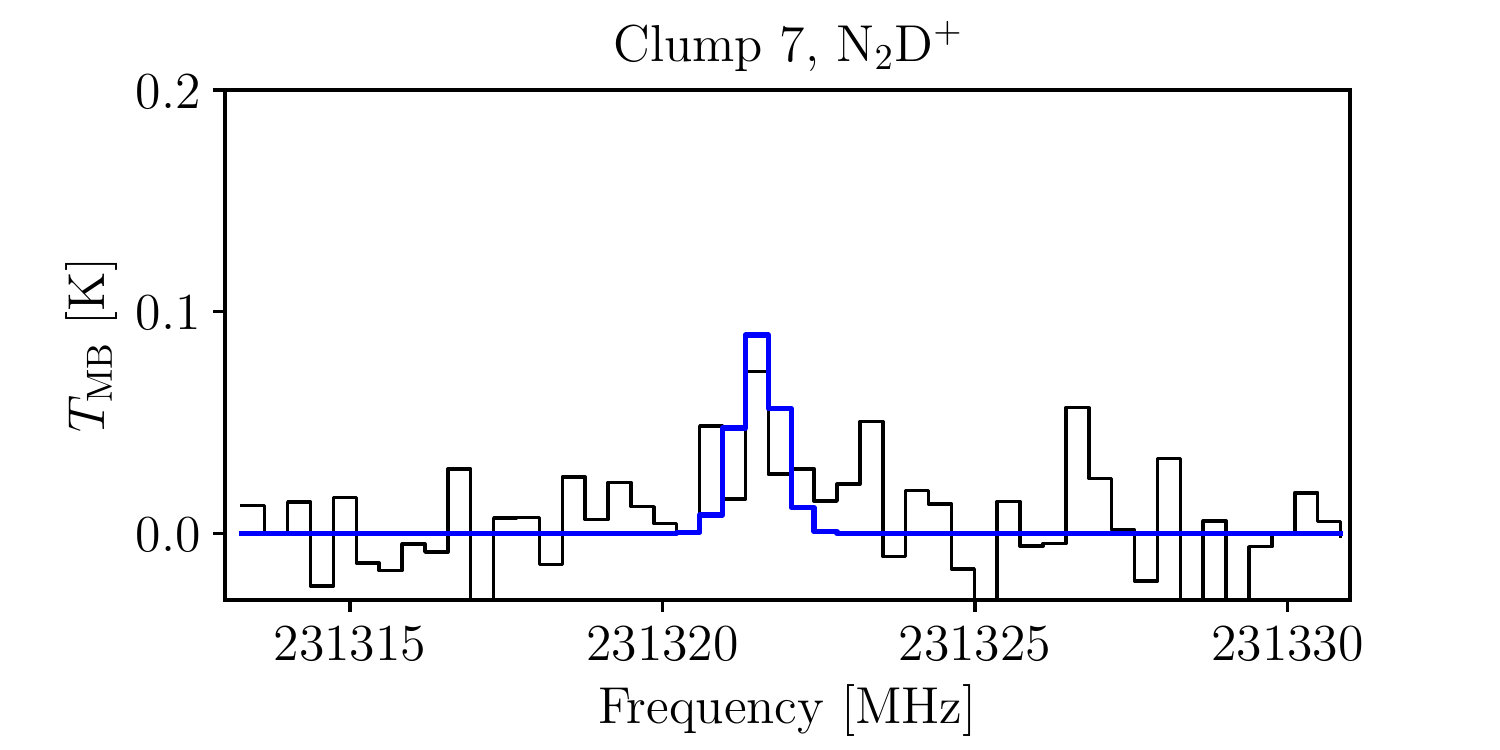}\\
        \includegraphics[width=0.5\textwidth]{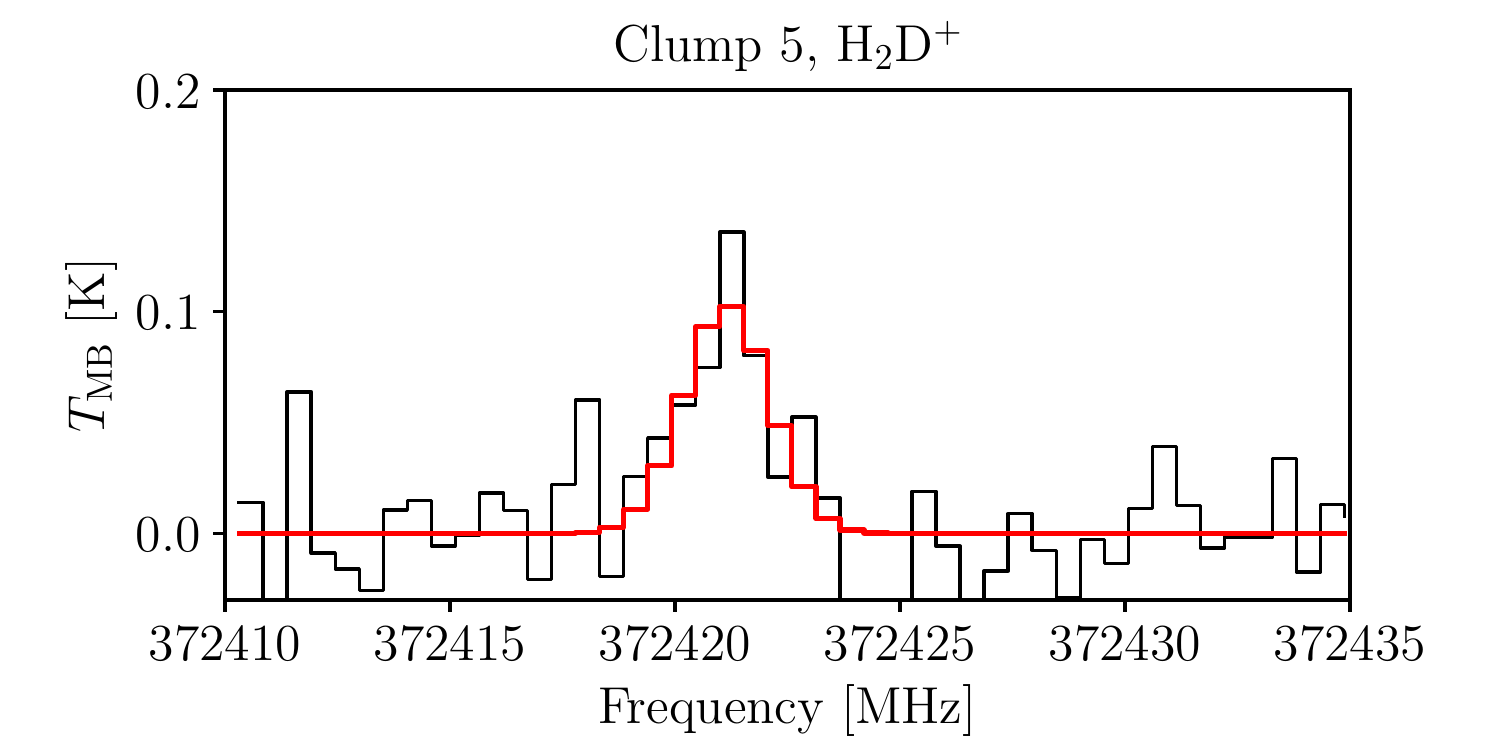}
        \includegraphics[width=0.5\textwidth]{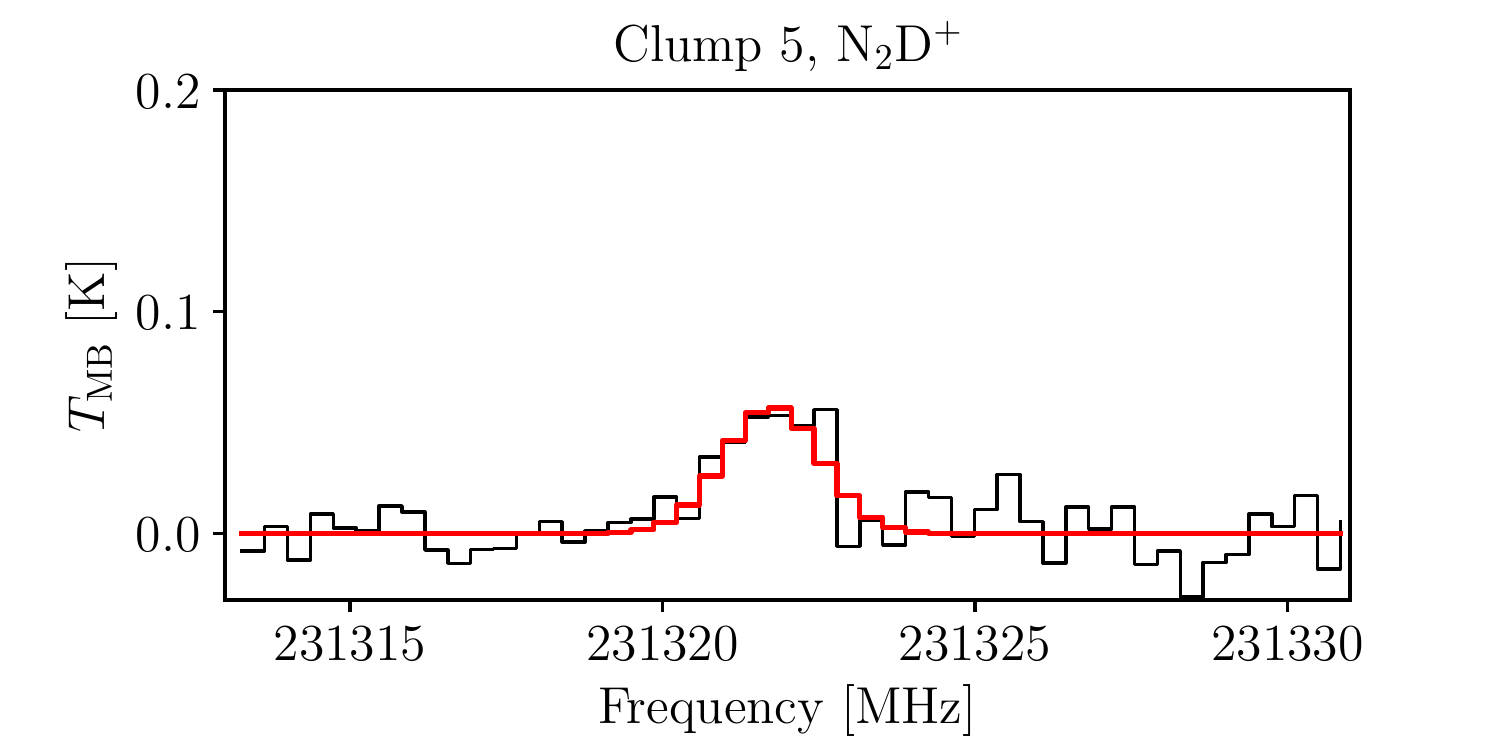}\\
        \includegraphics[width=0.5\textwidth]{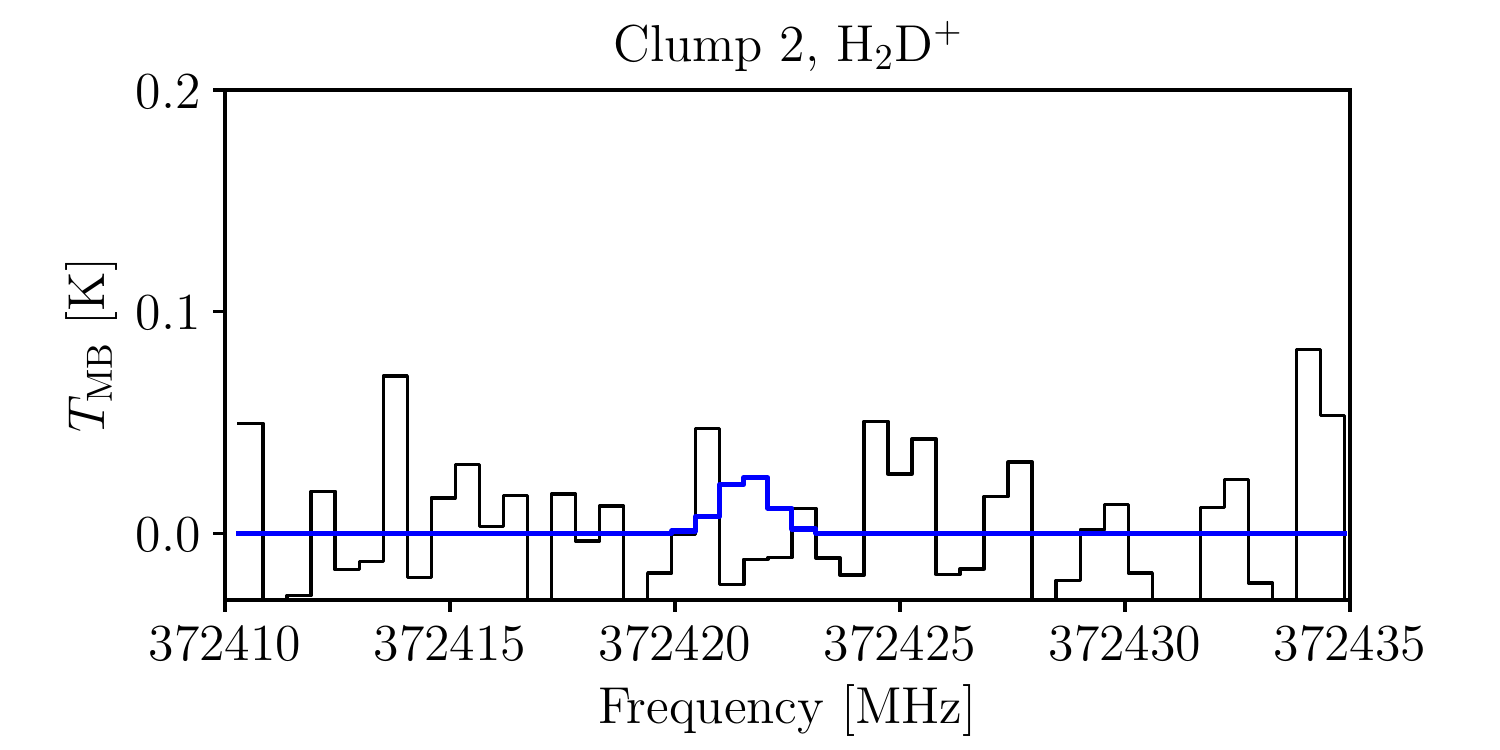}
        \includegraphics[width=0.5\textwidth]{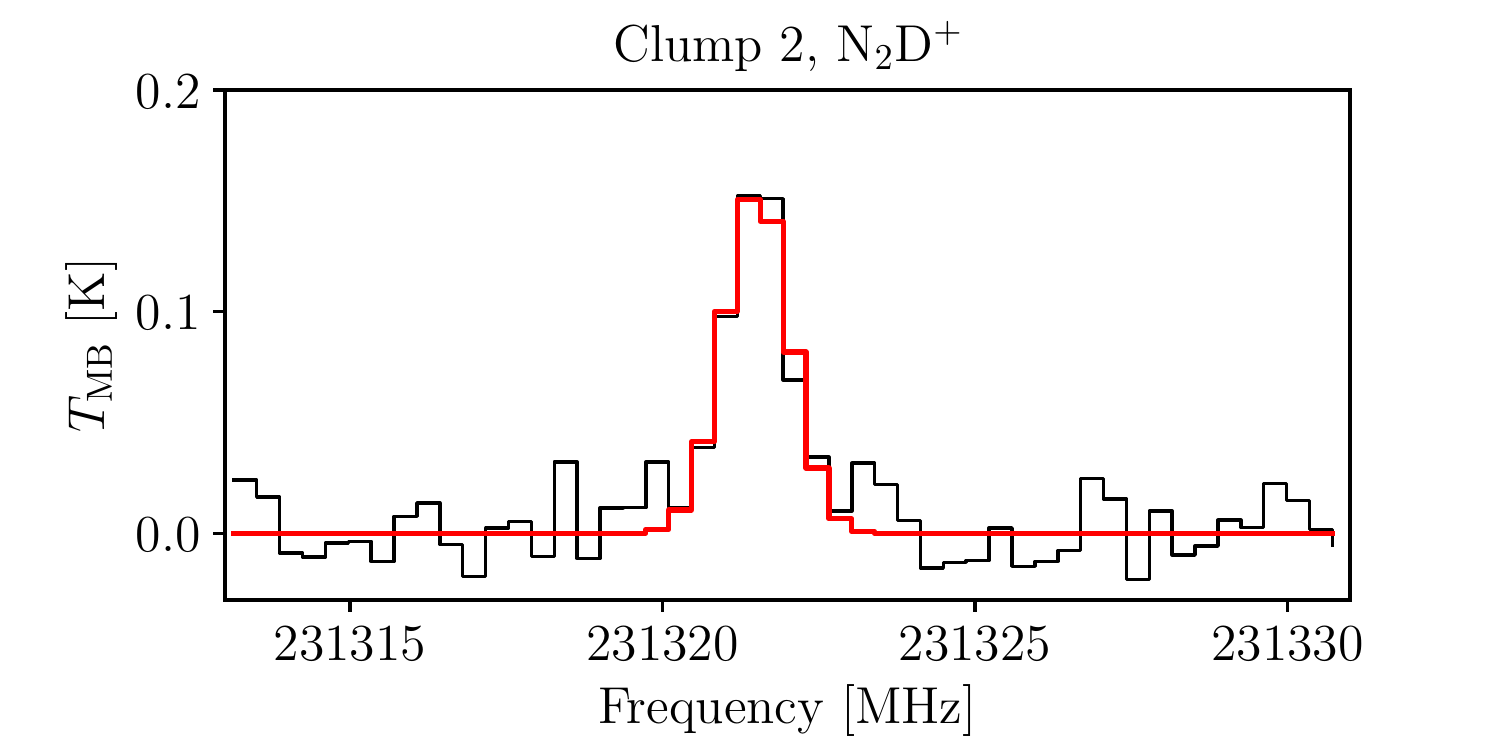}\\
        \caption{Spectra of the three clumps for o-\hhdp$(1_{1,0} - 1_{1,1})$ (left column) and \nndp(3--2) (right column). The best fit is indicated in red, while upper limits are drawn in blue.}\label{fig:spectra}
    \end{figure*}

    In our APEX observations Clumps 5 and 7 were detected in o-\hhdp$(1_{1,0} - 1_{1,1})$, while Clumps 2 and 5 were detected in \nndp(3--2). Figure~\ref{fig:spectra} shows the spectra for all sources observed, highlighting their different features in line emission.
    
    To give a physical explanation of this behaviour, we first characterise the properties of the clumps.
    
    \subsection{Dust spectrum}\label{sec:dust_spectrum}
    
    We perform aperture photometry on the clumps to extract dust continuum fluxes. We use images from ATLASGAL \citep[$870\mum$][]{Schuller+09_aap504_415}, HiGAL \citep[$350\mum$, $250\mum$, $160\mum$ and $70\mum$][]{Molinari+10_pasp122_314}, MIPSGAL \citep[$24\mum$][]{Carey+09_pasp121_76} and MSX \citep{Egan+03_AFRL}. All images were smoothed to $28\arcsec$, the resolution of the APEX 230 GHz observations, and the fluxes were extracted in the central beam, after removing the median background estimated in an annulus with $2 \times \theta_{beam} < R_{annulus} < 5 \times \theta_{beam}$, for each clump. The apertures and annuli are centred on the positions of the clumps, as determined in \citet{Leurini+11_aap533_85}.
    
    The dust temperature and column density are estimated via a greybody fit at $\lambda\geq70\mum$. The bolometric luminosity is computed as the sum of the integral of the best-fit greybody curve and the integral at $\lambda\leq70\mum$, computed with the trapezoidal rule in log-log space.
    For the greybody we use $\kappa_{870\mum}=1.85\cm^2\usk\gram^{-1}$ and $\beta=1.75$, and consider the $70\mum$ flux as an upper limit.
    
    To convert the dust column density to that of molecular hydrogen, we use a gas-to-dust ratio $\gamma=120$, as calculated from Eq.~2 in \citet{Giannetti+17_aap606_12}, and $R_{GC} = 7.4\kpc$ \citepLeurini
    . The impression that Clump 7 is the least evolved of the three sources considered, and that Clump 2 is the most evolved one is confirmed by $\td$ and the $L/M$ ratio (see Table~\ref{tab:clump_props}), two efficient indicators of evolution \citep[e.g.][]{Saraceno+96_aap309_827, Molinari+08_aap481_345, Koenig+17_aap599_139, Urquhart+18_mnras473_1059}.
    
    The average volume density along the line of sight listed in Table~\ref{tab:clump_props} is estimated as $n_{\mathrm{H_{2}}} = N_{\mathrm{H_{2}}} / \ell$, where $\ell$ is the size of the clump obtained by \citet{Leurini+11_aap533_85}. We note however that the filament width is nearly constant ($\sim0.2\pc$; \citealtLeurini), even at the location of the clumps, so the difference in mean volume density may be overestimated. 

    \begin{table*}
        \centering
        \caption{Properties of the clumps derived from dust continuum emission. The clumps go from the least- to the most evolved.}\label{tab:clump_props}
        \begin{tabular}{lrrrrrrr}
            \hline
            \hline
            Source  & $\td$ & $N(\mathrm{H_2})$ & $L_{bol}(R<14\arcsec)\tablefootmark{a}$ & $M(R<14\arcsec)\tablefootmark{a}$ & $M\tablefootmark{b}$ & $Diameter\tablefootmark{c}$ & $n(\mol{H_2})$ \\
            & K     & $\pot{22}\cm^{-2}$& $\lsuntab$         & $\msuntab$       &$\msuntab$& pc              & $\pot{5}\cm^{-3}$\\
            \hline
            Clump 7 &  13.0 &           7.3 &                      23 &                31 &          120 &            0.18 &        1.3 \\
            Clump 5 &  15.5 &           8.7 &                     111 &                35 &          100 &            0.10 &        2.9 \\
            Clump 2 &  20.0 &          10.5 &                     331 &                44 &          200 &            0.19 &        1.8 \\
            \hline
        \end{tabular}
        \tablefoot{\tablefoottext{a}{Within the central $28\arcsec$, see text.} \tablefoottext{b}{Rescaled for the new dust temperatures from the values in \citet{Leurini+11_aap533_85}.} \tablefoottext{c}{From \citet{Leurini+11_aap533_85}.}}
    \end{table*}
    
    \subsection{Molecular column densities and abundances}\label{sec:Nmol}

    To calculate the column densities, a spectral-line fit is performed, under the assumption of local thermodynamic equilibrium (LTE) \citep[cf.][]{Vastel+12_aap547_33}, with MCWeeds \citep{Giannetti+17_aap603_33}. This allows the uncertainty on this parameter to be obtained as well (Table~\ref{tab:results}). We use the partition function for o-\hhdp\ (in the relevant temperature range, $9.375 \kel$: $10.3375$, $18.750 \kel$: $12.5068$, $37.500 \kel$: $15.5054$) from CDMS \citep{Mueller+01_aap370_49}. 
    Because the observed clumps are sufficiently dense to attain thermal coupling between gas and dust (cf. Table~\ref{tab:clump_props}), and because we lack an estimate of the excitation temperature for o-\hhdp$(1_{1,0} - 1_{1,1})$ and \nndp(3--2), we use $\td$, assuming that $\td=\tex$. 
    The fit is performed using Monte Carlo Markov Chains. We used an adaptive Metropolis-Hastings sampler \citep{Haario+01_Bernoulli7_223}, with 100,000 total samples, a burn-in period and a delay for the adaptive sampling of 10,000 and 5,000 iterations, respectively, and a thinning factor of 20 \citep[see][for more details on these parameters]{Giannetti+17_aap603_33}. Convergence and independence of the samples are ensured with the Raftery-Lewis \citep{RafteryLewis95_PracticalMCMC_115}, Geweke \citep{Geweke92_BS4_169} and Gelman-Rubin \citep{GelmanRubin92_BS4_625} tests.
    
    We include in the budget a Gaussian calibration uncertainty with $\sigma = 5\%$. 
    Multiple tests have been performed on the priors to make sure that the choice of the latter is not crucial for the results.

\begin{figure*}
        \includegraphics[width=\columnwidth]{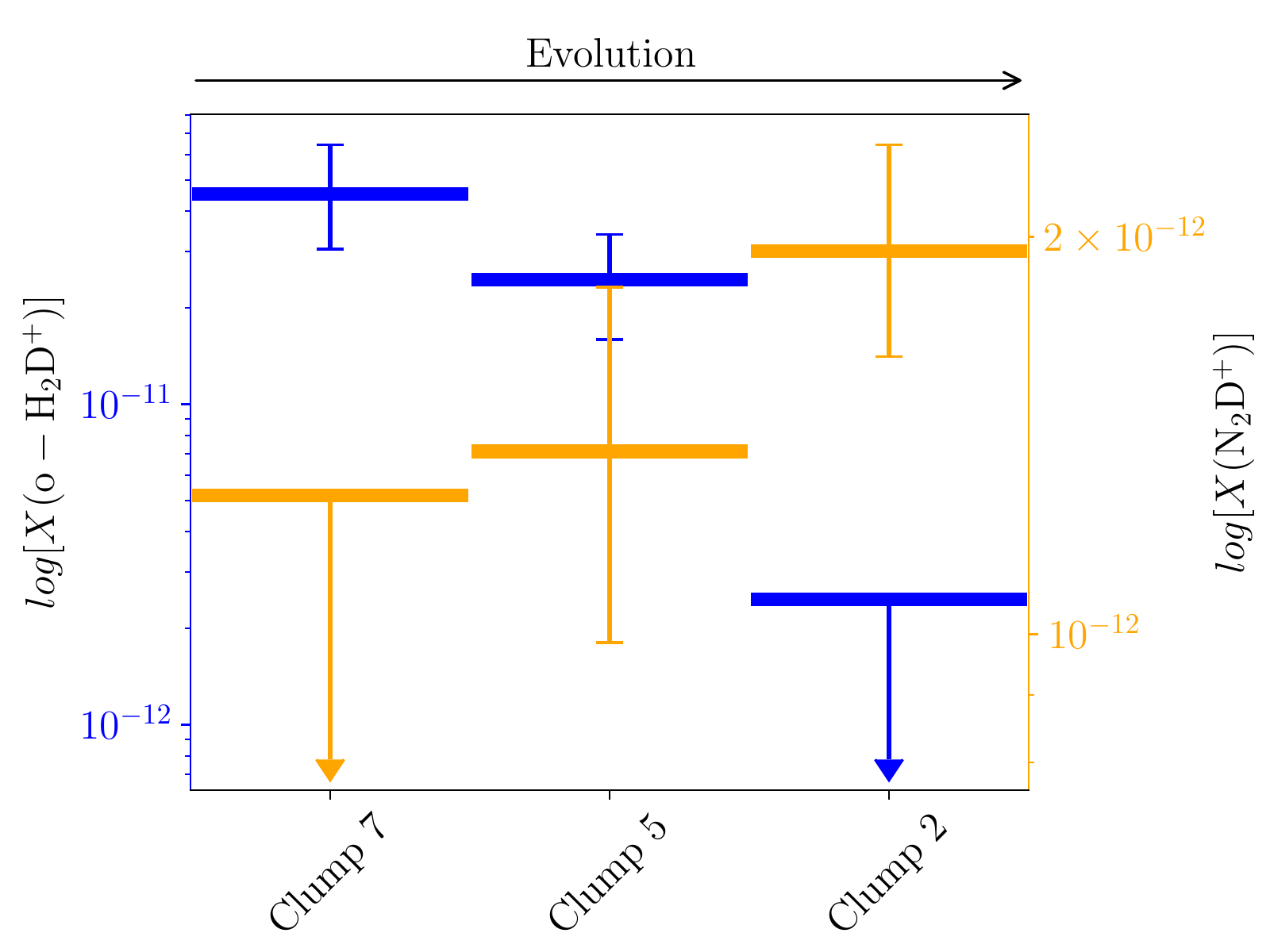}
        \includegraphics[width=\columnwidth]{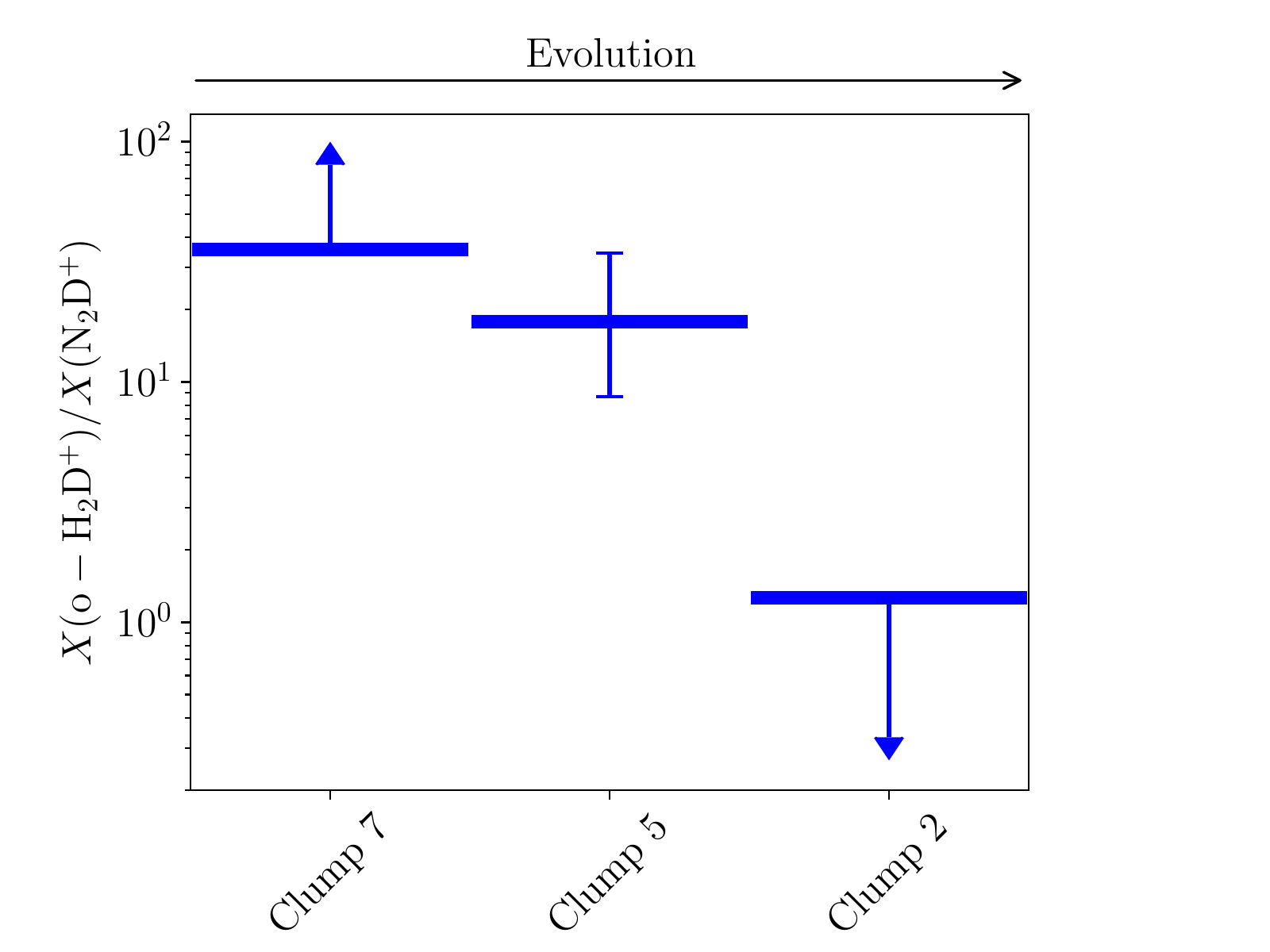}
        \caption{Left: Calculated o-\hhdp\ (blue) and \nndp\ (orange) abundances. Right: Relative abundance of the two species. The clumps are ordered by evolutionary stage, from the least- to the most evolved. The 95\% credible interval is indicated for abundances and ratios; we point out that for non-detections, only the upper limit is shown.}\label{fig:abundances}
    \end{figure*}

    Beam-averaged abundances can be estimated from the peak column density of H$_{2}$ derived from the dust continuum emission. In the following, the same value of $N_{\mathrm{H_{2}}}$ is used for both o-\hhdp\ and \nndp, despite the difference in the angular resolution of the spectral-line observations ($18-28\arcsec$, respectively). 
    Smoothing the continuum maps to the resolution of the ATLASGAL data (excluding the HiGAL $350\mum$ image) leads to an increase in the peak H$_{2}$ column density of $\lesssim15\%$, well within the uncertainties; this has the effect of lowering $X(\hhdp)$ in Clump 5, thus increasing the difference with Clump 7 (see Fig.~\ref{fig:abundances}, left).
    
    If $X(\mathrm{o-}\hhdp)$ does not vary significantly over scales of 18000-28000~AU ($18-28\arcsec$ at $1\kpc$), one can also estimate the relative abundance of the two species. 
    In Fig.~\ref{fig:abundances} one can see that o-\hhdp\ becomes rarer with evolution, while \nndp\ shows the opposite behaviour. Considering the abundance variation and its uncertainty, o-\hhdp\ is much more sensitive to the clump evolution than \nndp. The relative abundance (panel (b)) of the two species offers another way of looking at the same finding, but with the advantage of being independent of the H$_2$ column density, then removing an additional source of uncertainty. From the figure, we see that the o-\hhdp/\nndp\ ratio progressively decreases in time and o-\hhdp\ is more abundant than \nndp\ by a factor $\gtrsim10$ in the youngest Clump 5 and 7. 
        
    \begin{table*}
        \centering
        \small
        \caption{Results of the line-fitting procedure with MCWeeds.}\label{tab:results}
        \begin{tabular}{lrrrrrrrrrrrr}
            \hline
            \hline
            Source & \multicolumn{2}{c}{$N(\text{o-}\hhdp)$} & \multicolumn{2}{c}{$N(\nndp)$} & \multicolumn{2}{c}{$\vlsr(\text{o-}\hhdp)$} & \multicolumn{2}{c}{$\vlsr(\nndp)$} & \multicolumn{2}{c}{$\Delta V(\text{o-}\hhdp)$} & \multicolumn{2}{c}{$\Delta V(\nndp)$} \\
            & \multicolumn{2}{c}{$\pot{11}\cm^{-2}$} & \multicolumn{2}{c}{$\pot{10}\cm^{-2}$} & \multicolumn{2}{c}{$\kmstab$} & \multicolumn{2}{c}{$\kmstab$} & \multicolumn{2}{c}{$\kmstab$} & \multicolumn{2}{c}{$\kmstab$} \\
            & med. & $95\%$ CI & med. & $95\%$ CI & med. & $95\%$ CI & med. & $95\%$ CI & med. & $95\%$ CI & med. & $95\%$ CI \\
            \hline
            Clump 2 &  $\dots$ &      $< 2.6$ &     20.4 &  $17.0;24.6$ & --3.0 &      $\dots$ & --2.4 &  $-2.5;-2.3$ & 1.0 &    $\dots$ & 1.6 &  $1.3;1.9$ \\
            Clump 5 &     21.4 &  $13.9;29.6$ &     12.0 &   $8.6;16.0$ & --2.6 &  $-2.9;-2.3$ & --2.7 &  $-3.0;-2.3$ & 1.8 &  $1.1;2.6$ & 2.3 &  $1.7;3.2$ \\
            Clump 7 &     33.3 &  $22.4;47.2$ &  $\dots$ &      $< 9.3$ & --3.8 &  $-4.0;-3.5$ & --3.0 &      $\dots$ & 1.5 &  $1.0;2.6$ & 1.0 &    $\dots$ \\
            \hline
        \end{tabular}
        \tablefoot{For each quantity, we report the median value and the 95\% credible interval. 
        }
    \end{table*}

    \section{Discussion}\label{sec:discussion}
    
    In the previous section we show that o-\hhdp\ and \nndp\ have opposite trends in abundance as a function of evolution. In the following we discuss a few possible explanations for the abundance differences between the three clumps.
    
    \subsection{Deuteration of H$_3^+$ and availability of gas-phase N$_2$}
    
    \hhdp\ rapidly forms in cold and dense gas, where the reaction
    \begin{equation}
    \mathrm{H_{3}^{+}} + \mathrm{HD} \leftrightarrows \hhdp + \mathrm{H_{2}} + \Delta E\label{react:hhdp_formation}
    \end{equation}
    only proceeds in the forward direction, unless there is a substantial fraction of ortho-\mol{H_2} \citep[e.g.][]{Gerlich+02_planss50_1275}, the only case in which $\Delta E$ can be negative. \hhdp\ then drives the deuteration of the neutral molecules that have a deuteron affinity larger than molecular hydrogen \citep{Millar+89_apj340_906}.
    The o-\hhdp\ abundance derived in Clumps 5 and 7 is of the order of $\sim 3\times \pot{-11}$, similar to the values reported by \citet{Pillai+12_apj751_135} for their o-\hhdp\ peaks in the DR 21 region, and by \citet{Vastel+06_apj645_1198} for the outskirts of L1544.
    For the temperatures of our sources, $13\kel$ and $15.5\kel$ for Clumps 7 and 5, respectively, the ortho-to-para ratio (OPR) predicted by \citet{Flower+04_aap427_887} for \hhdp\ is in the range $0.05-0.1$. 
    Recently, we also performed 3D magnetohydrodynamic simulations of collapsing clumps and filaments, which included chemistry under the assumption of total depletion \citep{Koertgen+17_mnras469_2602, Koertgen+18_mnras478_95}. In these simulations we also follow the OPR for \hhdp, finding that it rapidly decreases to values below unity, in agreement with \citet{Flower+04_aap427_887}, on time scales of the order of $\pot{5}\yr$.
    This OPR can be used to estimate the total \hhdp\ abundance, which would be in the range $\pot{-10}-\pot{-9}$. \citet{Caselli+03_aap403_37} show how in L1544, \hhdp\ is nearly as abundant as electrons ($\pot{-9} \vs 2\times\pot{-9}$), indicating that this species is a major ion, which only happens when the depletion is extremely high. Our results, when compared to these findings, also suggest that regions exist in the clump where heavy elements are vastly depleted, which is also supported by our ongoing study of CO depletion for the entire filament \citepSabatini.
    
    Clump 7, however, not only has a high abundance of o-\hhdp, but also shows low values of $X(\nndp)$. One possible explanation for this is the time lag that is needed to form \nndp\ from \hhdp. As an example, Figure~1 in \citet{Sipila+15_aap578_55} shows that in the first $\pot{5}\yr$, according to their model, o-\hhdp\ is much more abundant than \nndp, and the difference progressively decreases. This provides, qualitatively, an upper limit to the age of the clump of 10$^5$ years, comparable to the current free-fall time ($t_{ff}\sim 1.2\times 10^5$ yr). 
    The clump is indeed sub-virial ($\alpha \sim 0.4$\footnote{Assuming a homogeneous clump \citep[see][]{MacLaren88}, the radius from Table~\ref{tab:clump_props}, and the line width from o-\hhdp$(1_{1,0} - 1_{1,1})$.}), and shows a very weak stellar activity in its centre. Clump 7 is also the only one to show a blue-skewed profile in \mol{HCO^+}(3--2), which indicates that the clump is collapsing (see Fig.~\ref{fig:hcop_spectra}).
    We note that if the \mol{H_2} ortho-to-para conversion occurs on the surface of dust grains, the chemical timescale could be even shorter \citep{Bovino+17_apjl849_25}.
    
    In Clump 5, on the other hand, $X(\nndp)$ is two to three times larger than in Clump 7. In this clump, being more evolved, \nndp\ has had time to form, while the stellar activity is not yet sufficient to warm up the gas 
    to temperatures high enough to significantly alter the chemistry on large scales. 
    
    Interestingly, Clump 2 still has a large fraction of \nndp\ with respect to \mol{H_{2}}, but at the same time $X(\hhdp)$ is at least an order of magnitude lower than $X(\nndp)$. Because \nndp\ forms via deuteron exchange from \hhdp, one would expect to see a correlation in the abundance of the two molecules, after they have had time to form. Furthermore, Clump 2 hosts luminous young stellar objects (YSOs), possibly close to the zero-age main sequence (ZAMS), as indicated by its $L/M$ ratio \citep{Giannetti+17_aap603_33}, and is therefore a relatively evolved source. 
    
    What is keeping the level of \nndp\ high, even higher than in Clump 5, and at the same time destroying o-\hhdp?
    The beam- and line-of-sight-averaged depletion of CO does not vary significantly from Clump 7 to Clump 2, and a relatively high depletion is one of the key ingredients for the efficient formation of \nndp\ because carbon monoxide acts as a destroyer of both \hhdp\ and \nndp\ \citep[e.g.][]{Walmsley+04_aap418_1035, Emprechtinger+09_aap493_89}.
    The other fundamental ingredient for the formation of \nndp\ is the presence of \mol{N_{2}} in the gas phase, which can react with \hhdp, \mol{D_2H^+}, and \mol{D_3^+}.
    Close to the YSOs, where the temperature is high and CO is in the gas phase, \hhdp\ and \nndp\ are both actively destroyed
    and Eq.~\ref{react:hhdp_formation} is also fast in the backward direction.
    However, at larger distances from luminous YSOs, the material is still dense and cold, and
    CO remains frozen out onto dust grains, contrary to \mol{N_{2}}, either because the latter has not been significantly depleted \citep[due to its longer formation timescale compared to CO and the fact that atomic nitrogen could have a lower sticking coefficient than \mol{N_2}][]{Flower+06_aap456_215}, or because it evaporates faster.
    In fact, according to the most recent estimates \citep{Wakelam+17_MolecularAstrophysics6_22} \mol{N_{2}} and CO binding energies differ by a few hundred Kelvin (1100 K vs 1300 K), enough to cause a significant difference in the evaporation timescales.
    The latter, evaluated by employing the standard $\tau_\mathrm{evap} = \nu_{0}^{-1} \mathrm{e}^{E_D / k_{B} T_\mathrm{d}}$, are $\sim 5\times\pot{8}\yr$, and $\sim 2\times\pot{4}\yr$, for CO and N$_2$, respectively.
    In the equation above, $\nu_{0}=\pot{12}\second^{-1}$ is the typical harmonic frequency \citep{Hasegawa+92_apjs82_167}, $E_D$ is the binding energy, $k_{B}$ is the Boltzmann's constant, and $T_\mathrm{d}$ is the dust temperature (see Table~\ref{tab:clump_props}).
    More importantly, with time, a large amount of \hhdp\ is converted to \mol{D_2H^+} and \mol{D_3^+}, decreasing its abundance. \nndp, on the other hand, is more efficiently formed by the reaction of \mol{D_2H^+} and \mol{D_3^+} with \mol{N_2}, compared to \hhdp.
    The combination of these processes could succeed in reproducing the observed reduction in abundance of o-\hhdp, and in maintaining a high column density of \nndp.
    In even more evolved sources, $X(\nndp)$ decreases, as expected \citep{Fontani+15_aap575_87}.
    
    If this is the case, we propose that the combined abundances of o-\hhdp\ and \nndp\ represent a more efficient evolutionary indicator for the first stages of the high-mass star-formation process, compared to their individual values \citep[e.g.][]{Caselli+08_aap492_703, Fontani+15_aap575_87}, in the same way that they are both needed to trace the full reservoir of gas \citep{Pillai+12_apj751_135}.

    \subsection{Alternative explanations}

    One alternative possibility to explain the emission in Clump 7 could be that the region in the clump where heavy elements (including \mol{N_2}) are completely depleted is extended, strongly affecting the abundance measurement. Contamination in the beam and along the line-of-sight would explain the weak emission of high-density tracers towards this source \citep[see also][for \mol{N_{2}H^{+}} $J=1\rightarrow0$]{Leurini+11_aap533_85}. 
    
    The critical density of the o-\hhdp$(1_{1,0} - 1_{1,1})$ is $\sim\pot{5}\cm^{-3}$, roughly one order of magnitude lower than that of \nndp(3--2), and similar to the mean values of $n(\mol{H_2})$ reported in Table~\ref{tab:clump_props}. Although this is not always a direct indication that emission is coming from denser gas \citep[see e.g.][]{Kauffmann+17_aap605_5}, it is possible that in Clump 7 the region dense enough to excite \nndp(3--2) is smaller than in the other two clumps. 
    
    In both cases, the abundance of o-\hhdp\ relative to \nndp\ keeps its potential as an evolutionary indicator, on the one hand because if a region of complete depletion exists at the centre of the clump, the warm-up by the YSOs will progressively erode it, and on the other hand because density increases as a result of the clump collapse, before feedback becomes important at the clump scale and starts to dissipate it.
    
    For Clumps 2 and 5, the \nndp(3--2) emission is likely to arise in cold gas surrounding the star-forming cores, as observed by \citet{Fontani+09_aap499_233} for example. If o-\hhdp\ is abundant in these compact structures, its emission may be completely saturated and diluted in the APEX beam for Clump 2; o-\hhdp$(1_{1,0} - 1_{1,1})$ must also be weak in the lower-density gas enveloping these cores. This is in contrast with the observations of extended emission in DR 21, and those of L1544 \citep{Vastel+06_apj645_1198}, although the envelope of high-mass star forming regions could be warmer due to the stronger external illumination from close-by high-mass stars; this increases the ortho-to-para \mol{H_2} ratio, which suppresses D-fractionation \citep[e.g.][]{Kong+15_apj804_98}. 
    Additional high-resolution observations are needed to investigate these possibilities.

    \section{Summary and conclusions}
    
    We observed the o-\hhdp$(1_{1,0} - 1_{1,1})$ and \nndp(3--2) lines in three clumps along the spine of G351, the most massive filament within $1\kpc$ \citep{Leurini+11_aap533_85} in ATLASGAL. 
    These observations provide for the first time the possibility to investigate the variation with evolution of the abundances of o-\hhdp\ and \nndp\ in high-mass sources, from a clump that is still quiescent at $70\mum$, to one which hosts luminous YSOs, likely close to the ZAMS. The selected clumps not only belong to the same complex, ensuring the best conditions for a comparison of the chemistry, but also have a comparable peak column density of \mol{H_{2}}.
    
    The abundance of \nndp\ progressively increases with evolution, while an opposite trend is found for o-\hhdp. We propose that the chemical evolution of the clumps causes this behaviour, and that the relative abundance of these species is a good indicator of time evolution. 
    
    First, after the clump reaches a high-enough density to allow for an efficient CO depletion, \hhdp\ starts to form, but to attain a considerable amount of \nndp\ would require a longer time \citep[e.g.][]{Sipila+15_aap578_55}. Clump 7 still has $X(\nndp) \lesssim 9\times\pot{-13}$, which also suggests an age of no more than 10$^5$ yr, and a fast collapse time. As time passes, more \nndp\ forms via reaction~\ref{react:nndp_formation} in cold and dense gas. High degrees of deuteration are commonly observed in relatively young and massive clumps \citep[e.g.][]{Caselli+02_apj565_344, Fontani+11_aap529_7}. This is the situation of Clump 5, where both \hhdp\ and \nndp\ are abundant, despite the presence of YSOs (Fig.~\ref{fig:multilambda}), which are not yet luminous enough to warm up a significant portion of the gas in the clump. 
    In Clump 2, $X(\nndp)$ increases by a factor of $\sim 2$ and $X(\hhdp)$ decreases by a factor $\gtrsim10$ compared to Clump 5. Because this source is the most evolved, \mol{N_2} has had more time to form. Also, the line-of-sight and beam-averaged dust temperature in Clump 2 is $\sim20\kel$, not high enough to release CO into the gas phase on large scales (see \citealtSabatini), but sufficient to reduce the evaporation timescale of \mol{N_{2}} to $\sim 2\times\pot{4}\yr$. 
    Therefore, \mol{N_{2}} is abundant in the gas phase throughout the clump, whether or not it was depleted from the gas phase earlier.
    \hhdp\ may be progressively transformed into \mol{D_2H^+} and \mol{D_3^+}, reducing the abundance of \hhdp, and, at the same time, boosting the production of \nndp. This behaviour could be confirmed with observations of p-\mol{D_2H^+}. 
    
    A weak \nndp(3--2) line in Clump 7 may alternatively be caused by excitation effects, because it has a critical density an order of magnitude larger than the o-\hhdp\ line \citep[see also][]{Pillai+12_apj751_135} and larger than the mean $n(\mol{H_2})$ of the source. A similar effect could result from a large part of the gas being affected by complete depletion of heavy elements. In both cases the potential of the relative abundance of o-\hhdp\ and \nndp\ in tracing evolution remains unaltered, because density increases with time, and the region of complete depletion is progressively eroded by the warm-up caused by the YSOs. 
    
    Without high-resolution observations, however, we cannot exclude that \hhdp\ is also abundant in compact \nndp-emitting regions for sources like Clump 2. Here, o-\hhdp$(1_{1,0} - 1_{1,1})$ could be completely saturated, and dilution in the APEX beam could cause the non-detection of the line.

    \begin{acknowledgements}
        SB is financially supported by CONICYT Fondecyt Iniciaci\'on (project code 11170268),  CONICYT programa de Astronomia Fondo Quimal 2017 QUIMAL170001, and BASAL Centro de Astrofisica y Tecnologias Afines (CATA) AFB-17002. DRGS is financially supported by CONICYT Fondecyt regular (project code 1161247) and the international collaboration project PII20150171. AG thanks the Astronomy Department, University of Concepci\'on for having supported his visit during July 2018 (via Redes Internacionales project number REDI170093). B.K.~acknowledges funding from the German Science Foundation (DFG) within the Priority Programm "The Physics of the ISM" (SPP 1573) via the grant BA 3706/3-2.
        This work was partly supported by the Collaborative Research Council 956, sub-project A6, funded by the Deut\-sche For\-schungs\-ge\-mein\-schaft (DFG).
        This paper is based on data acquired with the Atacama Pathfinder EXperiment (APEX). APEX is a collaboration between the Max Planck Institute for Radioastronomy, the European Southern Observatory, and the Onsala Space Observatory. This research made use of Astropy, a community-developed core Python package for Astronomy \citep[][http://www.astropy.org]{astropy_2013}, of NASA’s Astrophysics Data System, and of Matplotlib \citep{Hunter_2007_matplotlib}.
        MCWeeds makes use of the PyMC package \citep{Patil+10_jstatsoft35_1}.
    \end{acknowledgements}

    \bibliographystyle{aa}
    \bibliography{biblio.bib}
    
    \appendix
    \section{Additional spectra}
    \begin{figure}
        \includegraphics[width=0.5\textwidth]{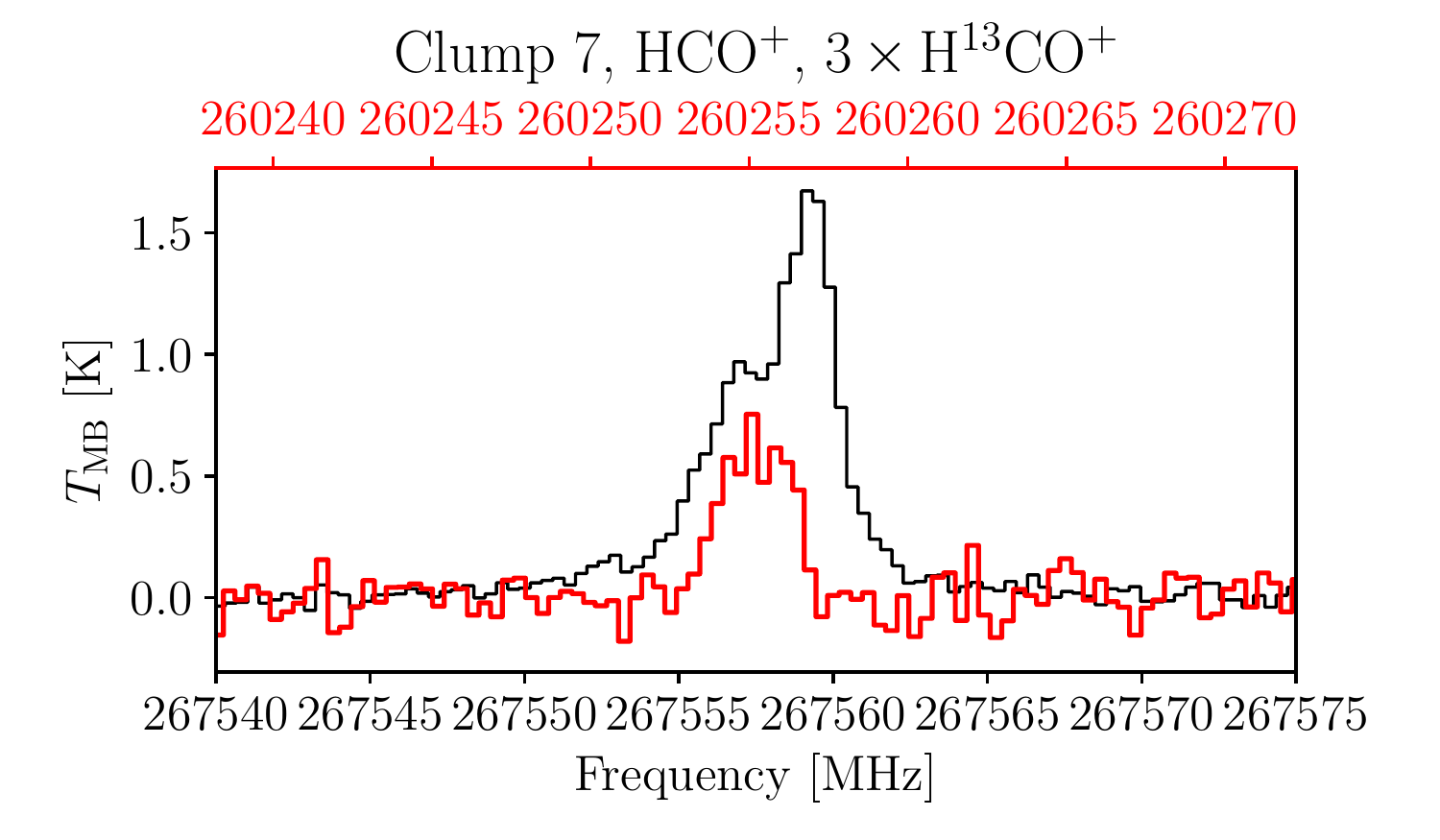}
        \includegraphics[width=0.5\textwidth]{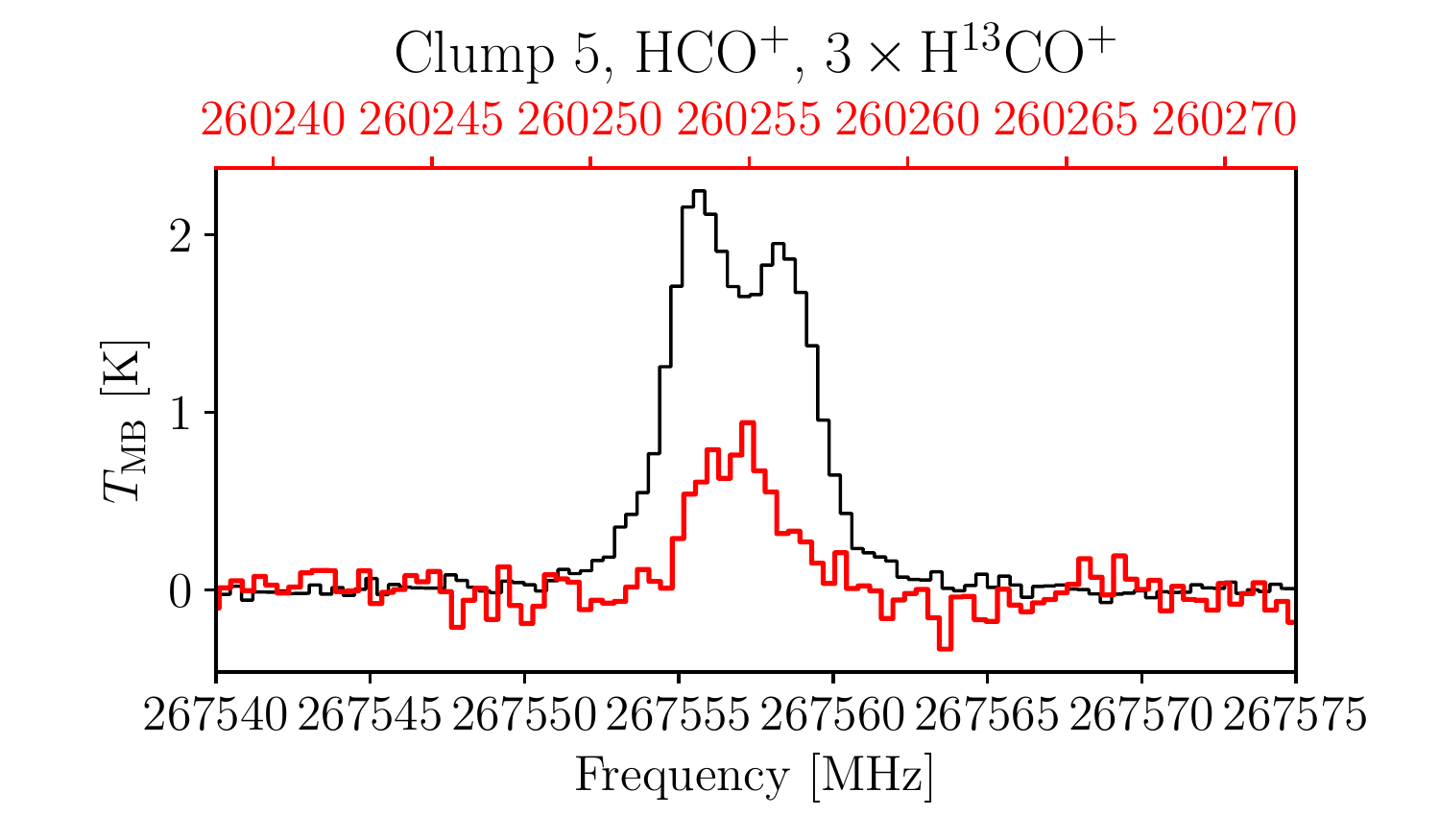}
        \includegraphics[width=0.5\textwidth]{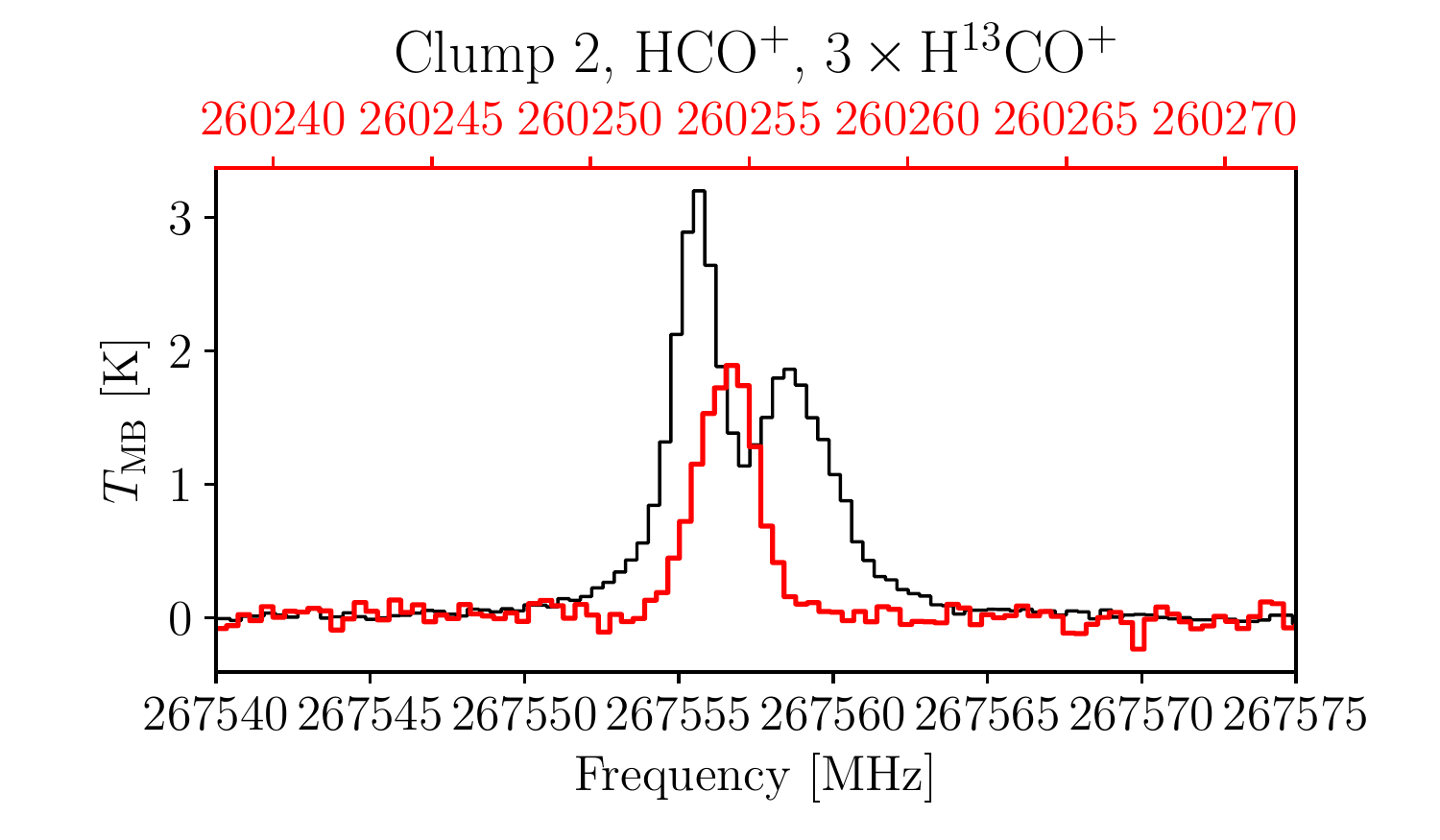}
        \caption{Spectra of the three clumps for \mol{HCO^+} $J=3\rightarrow2$ (black) and \mol{H^{13}CO^+} $J=3\rightarrow2$, multiplied by a factor of three (red).}\label{fig:hcop_spectra}
    \end{figure}
    
\end{document}